\documentclass[aps,groupedaddress,showpacs]{revtex4}
\usepackage{graphicx}
\def\bc{\begin{center}}
\def\ec{\end{center}}

\def\beq{\begin{equation}}
\def\eeq{\end{equation}}

\def\d{\downarrow}
\def\u{\uparrow}

\def\t{{\tau}}
\def\bj{{\bf j}}

\begin{document}

\title{Dimer states in atomic mixtures}

\author{K. Ziegler}
\affiliation{Institut f\"ur Physik, Universit\"at Augsburg\\
D-86135 Augsburg, Germany}
\date{\today}

\begin{abstract}
A mixture of heavy atoms in a Mott state and light spin-1/2 fermionic atoms 
is studied in an optical lattice. Inelastic scattering processes between both atomic species 
excite the heavy atoms and renormalize the tunneling rate as well as the interaction of 
the light atoms. An effective Hamiltonian for the latter is derived that describes
tunneling of single fermions, tunneling of fermionic pairs and an exchange of 
fermionic spins. Low energy states of this Hamiltonian are a N\'eel state for strong effective
repulsion, dimer states for moderate interaction, and a density wave of paired fermions
for strong effective attraction.
\end{abstract}


\maketitle

\section{Introduction}

Recent experimental work on ultracold bosonic and fermionic atoms opened a field to study
quantum states in optical lattices, like superfluid or Mott 
states \cite{bloch05,greiner02}, where the interaction between the atoms can be controlled  
by a magnetic field via a Feshbach resonance \cite{feshb}.  Spin-dependent effects 
\cite{ketterle03}, formation of dimers from fermionic atoms 
\cite{hulet03}, and mixtures of two atomic species 
\cite{stan04,esslinger06,sengstock06} provide opportunities for more complex quantum states.
An interesting example of the latter are dimer states. They have been discussed 
in solid-state systems
\cite{rokhsar88} and recently also for an ultracold Bose gas with ring exchange \cite{xu06}.
Frustrated spin systems were discussed for a number of atomic systems \cite{lewenstein06}.

The idea of this paper is to study a system in an optical lattice that undergoes
a first-order phase transition when a model parameter is changed. If this parameter 
is easily accessible in the experiment it can be adjusted exactly at the transition point.
As a consequence, the quantum system at the transition point describes a physical situation
that has a number of interesting features. First, there is the coexistence of the phases
in form of a metastable state,
leading possibly to phase separation. Second, new translational-invariant quantum states 
can appear. A system that realizes this type of physics is an atomic mixture which
is subject to ineleastic scattering between the atomic species. It can be controlled
by adjusting physical parameters, including the optical-lattice parameters (by
choosing the frequency and amplitude of the Laser field) and the fermion-fermion
interaction through a Feshbach resonance. This enables us to create ground states as
well as excited states of the system, and to study their dynamics. 
  
In the specific case of this paper,
an optical lattice is filled with heavy (bosonic or fermionic) atoms (HA)
(e.g. $^{87}$Rb or $^{40}$K), one at each lattice well, which form a Mott state.
The tunneling of these HA is neglected since the potential
barriers of the optical lattice are sufficiently high. Excitations are only due to
collisions with other atoms. For this purpose light fermionic atoms 
(LFA) (e.g. $^6$Li) are added to the optical lattice. 
These atoms can tunnel because of their low mass
and scatter by the HA. It is assumed that the HA experience a
harmonic potential in the optical lattice, at least at low energies. 
Then their excitations are harmonic-oscillator states. During the
scattering process the HA can also transfer energy to the LFA.
Moreover, we consider spin-1/2 fermions with  local (on-site) repulsion.
This can be described by a Bose-Fermi model that is
known in solid-state physics as the Holstein-Hubbard model \cite{alexandrov95}. 
The excitations of the HA are local phonons. 
This model is quite different from models
with density-density interaction, where only thermal fluctuations
of the HA and elastic scattering between the atomic species were studied \cite{ates05}.

\section{Holstein-Hubbard Model}

LFA with spin $\sigma$ ($=\u,\d$), defined by fermionic 
creation and annihilation operators $c_{\bj\sigma}^\dagger$ and  $c_{\bj\sigma}$
in a well of the optical lattice with position $\bj$, tunnel between nearest-neighbor 
wells. The corresponding tunneling Hamiltonian in 
tight-binding approximation reads
\beq
H_t=-J\sum_{<\bj,\bj'>}\sum_{\sigma=\u,\d}
(c_{\bj\sigma}^\dagger c_{\bj'\sigma}+h.c.) .
\label{h_t}
\eeq
HA, forming a Mott state,
are harmonic oscillators at each well with eigenfrequency $\omega_0$ and
energies $E_N=\omega_0N$ ($N=0,1,...$).
It is assumed that a HA in one well is excited independently of the HA in the other wells.
This gives the Hamiltonian $\omega_0 \sum_{\bj}b_\bj^\dagger b_\bj$,  where
$b_\bj^\dagger$ and $b_\bj$ are
creation and annihilation operators for the oscillator excitations
at each well, respectively. They can be considered as local phonons. 
LFA are scattered by HA due to an exchange 
of excitations of the HA (i.e. by creating or absorbing an excitation),
defined by the Hamiltonian
\beq
H_I=\sum_{\bj}\Big[\omega_0 b_\bj^\dagger b_\bj
+g(b_\bj^\dagger+b_\bj)(n_{\bj\u}+n_{\bj\d})
+Un_{\bj\u}n_{\bj\d}\Big] .
\label{h_0}
\eeq  
Here a local repulsive interaction between the LFA with strength $U\ge0$ has been included.
The phonons couple to the fermionic densities $n_{\bj\u}$ and $n_{\bj\d}$ with
strength $g$. Thus the atomic mixture is defined by the Hamiltonian $H=H_t+H_I$,
also known as the Holstein-Hubbard model \cite{alexandrov95}.

For a given ensemble of LFA, represented by integer
numbers $n_{\bj \sigma}=0,1$, the Hamiltonian $H_I$
can be diagonalized with product states
$
\prod_\bj |N_\bj;\sigma_\bj\rangle
$
(for $N_\bj=0,1,2,...; \sigma_\bj=0,\u,\d,\d\u$).
$|N_\bj;0\rangle$ is an eigenstate of the 
phonon-number operator $b_\bj^\dagger b_\bj$ with eigenvalue $N_\bj$.
A lattice product of these states is an eigenstate of the Hamiltonian $H_I$. 
Other eigenstates of $H_I$ with fermions can be constructed from
\beq
|N_\bj;\sigma\rangle=c_{{\bf j}\sigma}^\dagger
\exp[-{g\over\omega_0}(b_\bj^\dagger
-b_\bj)]|N_\bj;0\rangle\ \ \ (\sigma=\u,\d)
\label{single1}
\eeq
and from a state with two fermions per well
\beq
|N_\bj;\d\u\rangle=c_{{\bf j}\u}^\dagger
c_{{\bf j}\d}^\dagger
\exp[-2{g\over\omega_0}(b_\bj^\dagger -b_\bj)]|N_\bj;0\rangle.
\label{double}
\eeq
The eigenvalue of $\prod_\bj |N_\bj,\sigma_\bj\rangle$ with respect to $H_I$
is 
\beq
\sum_{\bj}\Big[\omega_0 N_\bj-{g^2\over\omega_0}(n_{\bj\u}+n_{\bj\d})^2
+Un_{\bj\u}n_{\bj\d}\Big] ,
\label{energy0}
\eeq
where the integers $n_{\bj\sigma}=0,1$ count the number of fermions at the well $\bj$
with spin $\sigma$. This expression
represents an effective {\it attractive} fermion-fermion interaction of strength
$g^2/\omega_0$ which competes with the {\it repulsive} fermion-fermion interaction
of strength $U$. Thus the system is given by the transformed Hamiltonian
\beq
H_\gamma =TH_tT^\dagger +\omega_0\sum_\bj b^\dagger_\bj b_\bj 
+\gamma\sum_{\bj}n_{\bj\u}n_{\bj\d}
-{g^2\over\omega_0}\sum_{\bj}(n_{\bj\u}+n_{\bj\d})
\label{newham}
\eeq
with the effective coupling $\gamma=U-2g^2/\omega_0$ and the unitary operator
\beq
T=\prod_\bj \left(\sum_{N_\bj\ge 0}
\sum_{\sigma=0,\u,\d,\d\u}e^{-\alpha_ng(b^\dagger_\bj-b_\bj)/\omega_0}
|N_\bj;\sigma\rangle\langle N_\bj;\sigma |\right) \ ,
\label{unitary}
\eeq
where $\alpha_0=0$, $\alpha_\u=\alpha_\d=1$, and $\alpha_{\d\u}=2$.
The states $|N_\bj;\sigma\rangle$ contain phonons even for $N_\bj=0$ due to the 
exponential factors in Eqs. (\ref{single1}) and (\ref{double}).
Therefore, $N_\bj>0$ will be called phononic excitation, in contrast to $N_\bj=0$,
which is the phononic ground state. The ground states of $H_I$ have $N_\bj=0$ and
are $\prod |0;\d\u\rangle$ for $U<3g^2/\omega_0$ and 
$\prod (a_1|0;\d\rangle + a_2 |0;\u\rangle)$ for $U>3g^2/\omega_0$
with $|a_1|^2+|a_2|^2=1$. The corresponding energies are $E_0=U-4g^2/\omega_0$
and $E_0=-g^2/\omega_0$, respectively.
For strong repulsive interaction there is only one fermion in each well, 
whereas a strong attractive interaction leads to a formation of a local fermion pair
(i.e. a bipolaron \cite{alexandrov95}). Both interactions are balanced for $U_c=3g^2/\omega_0$,
leading to a degeneration of the two states. 
The ground-state energy $E_0$ as a function of $U$ has a cusp at $U=U_c$.

It is crucial that unpolarized spin 1/2 fermions are considered. In the case of 
spin-polarized fermions (e.g. in a magnetic trap), the only local interaction
is due to Pauli's principle. This excludes multiple occupation of a well by
LFA. Consequently, the coupling to the HA results only in a $g$-dependent chemical
potential $\mu'=-g^2/\omega_0$ because of $n_\bj^2=n_\bj$ in Eq. (\ref{energy0}).
This favors the complete filling of the optical lattice with one LFA per well.

\subsection{Half-filled States}

Now it is assumed that the atomic system is prepared such that the number of LFA is
equal to the number of lattice wells (i.e. $2M$ fermions for $2M$ lattice wells).
This is the case of half filling and it is of particular interest because it allows 
highly degenerate quantum states. The energy of $TH_IT^\dagger$ for a half-filled system
is calculated as a state that can have $2j$ singly occupied wells with
energy $E_1=-2(g^2/\omega_0)j$ and $M-j$ doubly occupied wells with energy
$E_2=(U-4g^2/\omega_0)(M-j)$. The total energy then is 
\beq
E_0=\epsilon M-\gamma j{\ \ } {\rm with}{\ \ }
\epsilon=U-4g^2/\omega_0\ .
\label{ground0}
\eeq
Here $\gamma$, the effective fermion-fermion interaction of Eq. (\ref{newham}),
is an important parameter that controls
two different phases: For $\gamma>0$ the ground state has $j=M$ 
(i.e. $2M$ singly occupied wells), whereas for $\gamma<0$ the ground state has $j=0$ 
(i.e. $M$ doubly occupied and $M$ empty wells). The ground-state energy is 
$-2(g^2/\omega_0)M$ for $\gamma>0$ and $\epsilon M$ for $\gamma<0$, respectively. 
At $\gamma=0$ the ground-state energy $E_0$ does not depend on the number
of singly occupied wells, i.e. it is highly degenerate. 
This point describes a first-order phase transition in terms of $TH_IT^\dagger$, 
changing from a singly
occupied lattice for $\gamma>0$ to a mixture of doubly occupied and empty wells for $\gamma<0$.
Moreover, both states are highly degenerate even for $\gamma\ne0$
because the {\it local} Hamiltonian $H_I$ does not determine how the spins, the empty, and
the doubly occupied wells are distributed in the lattice.  At $U=U_c$
a pair of singly occupied wells can be
replaced by a pair of a doubly occupied and an empty well without costing
energy. These degeneracies are lifted by
the tunneling term $H_t$ if the more realistic Hamiltonian $H=H_t+H_I$ is considered.
It is known that this leads to a staggered spin state (N\'eel state) if $U\gg U_c$ and to 
a state with alternating empty and doubly occupied wells if $U\ll U_c$ \cite{alexandrov95}. 
The latter is a
density wave of fermion pairs. This is plausible if a nearest-neighbor pair of wells is 
considered: Two tunneling processes are possible in the case of a 
pair of singly occupied wells with opposite spin orientation but due to Pauli's principle 
no tunneling if the spin orientation is equal. Two tunneling processes are also possible in 
the case of a nearest-neighbor pair of an empty and a doubly occupied well. On the other hand,  
tunneling is blocked by Pauli's principle for a nearest-neighbor pair of doubly occupied
wells. Since tunneling lowers the energy,
the N\'eel state is close to the ground state for $U\gg U_c$ and the density wave is the 
ground state for $U\ll U_c$. In the remainder of this paper the situation of $U\approx U_c$
shall be studied.

\subsection{Partition Function and Resolvent}

Many physical quantities of an atomic system at temperature $T=1/\beta$ can be calculated
from the partition function
\beq
Z=Tr e^{-\beta H}
=\int_\Gamma Tr \left[(z-H)^{-1}\right]e^{-\beta z}{dz\over2\pi i}
\label{partition0}
\eeq
which is expressed here by the resolvent
$
G(z)=(z-H)^{-1}
$.
$\Gamma$ is a closed contour in the complex plane that encloses all eigenvalues of $H$ and 
$Tr$ is the trace over all quantum states of the system. At low temperatures this expression
is dominated by the lowest eigenvalues of $H$, and one can use the approximation
$Z\approx Z_0$ with
\beq
Z_0
=\int_\Gamma Tr \left[P_0(z-H)^{-1}P_0\right]e^{-\beta z}{dz\over2\pi i},
\label{partition1}
\eeq
where the trace is projected by the projector $P_0$ to quantum states that are 
energetically close to the ground state of the Hamiltonian $H$. In the specific
case of $H=H_t+H_I$ 
the restriction to states without phononic excitations (i.e. $N_\bj=0$)
\beq
P_0=\prod_\bj\Big(\sum_{\sigma_\bj}|0;\sigma_\bj\rangle\langle 0;\sigma_\bj |
\Big) 
\label{proj0}
\eeq
is an obvious choice
for sufficiently large $\omega_0$ (i.e. 
sufficiently strong amplitude of the optical lattice).
However, virtual phononic excitations are possible, excited by the Hamiltonian
of $e^{-\beta H}$ in the partition function. 

As an example, we consider the case without tunneling ($J=0$). The projection $P_0$
is onto half-filled states.
The poles of the projected resolvent then read
$z_0=\epsilon M$ for a density-wave state and $z_0=(\epsilon-\gamma) M$ 
for a single-fermion state. Depending on the parameter $\gamma$, the partition function
$Z_0$ is dominated by the lowest pole:
\[
-{1\over\beta M}\log(Z_0)=\epsilon -\gamma\Theta(\gamma) \ ,
\]
where $\Theta(\gamma)$ is the Heaviside function. At the transition point $\gamma=0$
this expression is continuous in $\gamma$ but its first derivative 
jumps with respect to $\gamma$.
According to the Ehrenfest classification scheme of phase transitions, 
this behavior characterizes a first-order phase transition.

\section{Recursive Projection Approach}

In order to evaluate the $P_0$--projected resolvent 
$
G_0(z)=P_0(z-H)^{-1}P_0
$
a recursive projection approach can be applied \cite{ziegler03}. The main idea of this method
is to control the mapping of states $|\Psi'\rangle=H|\Psi\rangle$, provided by the 
Hamiltonian $H$. For an initial state $|\Psi_0\rangle$ a sequence of linear independent
states $|\Psi_k\rangle$ ($k=1,2,...$) is created by a successive application of $H$.
This idea has been used in the Lanczos method and in the doorstep approach 
\cite{salgueiro05}. The sequence of states is
related to a sequence of projectors $P_k$, acting on the Hilbert space of all
quantum states of the system. For an initial projector $P_0$ and its complement
$P_1=1-P_0$ this sequence is created by the Hamiltonian $H$. For instance, projector 
$P_2$ is defined by applying $H$ to $P_0$ and projecting with $P_1$: $P_2$ is the projector
of the space that is created by the operator $P_{1}HP_{0}$.
This can be repeated by using $P_{3}HP_{2}$ to define $P_4$, where $P_3=P_1-P_2$ etc. 
It is important that the projectors with even index $P_{2j}$ 
project onto separate subspaces. An advantage of the method is that for a proper choice of
$P_0$ the spectrum of the projected Hamiltonians is shifted to higher energies at each
step. In the specific case of this paper the shift will be provided by phononic excitations. 
With the projections a sequence of projected resolvents is defined as
\[
G_0=P_0(z-H)^{-1}P_0 ,
\]
\[
G_{2j+2}=P_{2j+2}\left(P_{2j+1}(z-H)P_{2j+1}\right)^{-1}P_{2j+2}
\]
for $j=0,1...$ which are connected by the identity 
(more details in Ref. \cite{ziegler03})
\beq
G_{2j}=(z-P_{2j}HP_{2j} - P_{2j}HG_{2j+2}HP_{2j})^{-1} .
\label{projected3a}
\eeq
The iteration terminates at the $j^{th}$ level when $P_{2j}$ projects onto
a subspace whose basis consists only of eigenstates of $H$. Such a subspace we call eigenspace.
Moreover, when $P_{2j}$ projects on a subspace that is close to an eigenspace
(i.e. a basis of this subspace has very large overlap with eigenstates), the
truncation of the iteration is a good approximation. In terms of the Hamiltonian
this means that $HP_{2j}$ and $P_{2j}H$ are small.
This fact allows us to control the quality of the approximation.

Here only the lowest poles are of interest because they dominate the partition
function in Eq. (\ref{partition1}).
Depending on the Hamiltonian $H$ and the choice of the initial projection
$P_0$, the lowest pole $z_0^{j}$ of the projected resolvents $G_{2j}(z)$ can be 
increasing with $j$. If the values of $z$ are restricted to those where
$G_0(z)$ has poles but $G_2(z)$ has not, 
the sequence of projected resolvents can be truncated by approximating
$G_2(z)$. 
The recursive projection method, given by Eq. (\ref{projected3a}), is now 
employed for the Hamiltonian $H_\gamma$ of Eq. (\ref{newham}).
This Hamiltonian describes tunneling of LFA in the optical lattice and
interaction with HA, and direct interaction between the LFA.
 
The initial projector $P_0$ is given in Eq. (\ref{proj0}). This implies
\[
P_1=1-P_0={\sum_{\{N_\bj\ge0\}}}'
\prod_\bj\Big(\sum_{\sigma_\bj}|N_\bj;\sigma_\bj\rangle\langle N_\bj;\sigma_\bj |
\Big) ,
\]
where $\sum'$ restricts the summation to $\sum_\bj N_\bj\ge1$.
The $P_0$--projected Hamiltonian experiences a renormalization of the tunneling
rate: 
\[
P_0H_\gamma P_0=P_0\left[
TH_tT^\dagger+\omega_0\sum_\bj b^\dagger_\bj b_\bj +\gamma\sum_{\bj}n_{\bj\u}n_{\bj\d}
-{g^2\over\omega_0}\sum_{\bj}(n_{\bj\u}+n_{\bj\d})
\right]P_0
\]
\beq
=(\epsilon-\gamma) M+e^{-g^2/\omega_0^2} P_0H_t P_0 
+\gamma\sum_{\bj}P_0 n_{\bj\u}n_{\bj\d}P_0
\ .
\label{projham}
\eeq
An important consequence of the coupling between HA and LFA is the renormalization
of the tunneling rate as
\beq
J\to \tau=e^{-g^2/\omega_0^2}J \ ,
\label{ren}
\eeq
which is the well-known polaron effect \cite{alexandrov95}.

\subsection{Effective Hamiltonian}

The recursive projection will be applied to a restricted Hilbert space,
where the recursion terminates at level $j=2$. This should be sufficient
for small tunneling (i.e. for $J\ll \omega_0$). 
Then the resolvent $G_0$ can be written as $G_0=(z-H_0)^{-1}$ with the
effective Hamiltonian
\beq
H_{0}=P_{0}HP_{0} +P_{0}HG_{2}HP_{0} \ ,
\label{h00}
\eeq
and $G_2=(z-H_2)^{-1}$ with 
\beq
H_{2}=P_{2}H_IP_{2}+P_{2}H_tP_{2} +P_{2}HG_{4}HP_{2} \ . 
\eeq
The first term is
\beq
P_{2}H_IP_{2}=(\epsilon-\gamma) M+\omega_0\sum_{\bj}P_2b^\dagger_\bj b_\bj P_2
+\gamma \sum_{\bj}P_2 n_{\bj\u}n_{\bj\d}P_2 \ ,
\eeq
whereas the third term vanishes because $HP_{2}$ produces only states 
{\it outside} of the restricted Hilbert space. Moreover, it is assumed
that the second term vanishes because $H_tP_{2}$ produces only states
outside of the $P_2$-projected Hilbert space. This allows us to evaluate $G_2$
and subsequently the effective Hamiltonian $H_0$ in 
Eq. (\ref{h00}). The first part of $H_0$ was given in Eq. (\ref{projham}), 
and the second part $P_0HG_2HP_0$ must be calculated now. The latter separates 
into two parts, namely a term $H_s$ that exchanges single 
fermions with opposite spin and a term $H_p$ that exchanges a fermionic pair with an empty well:
\beq
H_s=\t^2\sum_{<\bj,\bj'>}\sum_{\sigma,\sigma'=\u,\d}
P_0c_{\bj \sigma}^\dagger c_{\bj'\sigma}
G_2
c_{\bj'\sigma'}^\dagger c_{\bj\sigma'}P_0
\equiv \sum_{<\bj,\bj'>}h_{1,\bj,\bj'}
\label{ham1}
\eeq
describes the exchange of spins: $h_{1,\bj,\bj'}$ gives 
$(\d,\u)\longleftrightarrow(\u,\d)$ for $\sigma\ne\sigma'$ and is diagonal
for $\sigma=\sigma'$. Moreover, 
\beq
H_p=\t^2\sum_{<\bj,\bj'>}\sum_{\sigma,\sigma'=\u,\d}
P_0c_{\bj \sigma}^\dagger c_{\bj'\sigma}
G_2
c_{\bj \sigma'}^\dagger c_{\bj'\sigma'}P_0\equiv \sum_{<\bj,\bj'>}h_{2,\bj,\bj'}
\label{ham2}
\eeq
describes the tunneling of two fermions as the exchange of a doubly occupied
well with an empty well $(0,\d\u)\longleftrightarrow(\d\u,0)$.

\subsection{Eigenstates for two wells}

Now the effective Hamiltonian $H_0$ for two wells is studied, where
\beq
H_0=\epsilon-\gamma 
+h_{t,\bj,\bj'}+h_{1,\bj,\bj'}+h_{2,\bj,\bj'}+\gamma(n_{\bj\u}n_{\bj\d}+n_{\bj'\u}n_{\bj'\d})
\label{hjj}
\eeq
with the single fermion tunneling term 
\[
h_{t,\bj,\bj'}
=-\tau\sum_\sigma\left( c^\dagger_{\bj\sigma}c_{\bj'\sigma}
+c^\dagger_{\bj'\sigma}c_{\bj\sigma}\right) \ .
\]
Eigenstates of $H_0$ on the $P_0$-projected Hilbert space can be constructed from dimer states:
\beq
|s_{\bj,\bj'}\rangle = (|\d,\u\rangle-|\u,\d\rangle)/\sqrt{2}\ ,
\hskip0.5cm
|d_{\bj,\bj'}\rangle = (|\d\u,0\rangle+|0,\d\u\rangle)/\sqrt{2}\ .
\label{dimers}
\eeq
These are eigenstates of $h_{1,\bj,\bj'}$ with eigenvalues $K_1$ and 0
\beq
h_{1,\bj,\bj'}|s_{\bj,\bj'}\rangle =K_1|s_{\bj,\bj'}\rangle \ ,
\hskip0.5cm
h_{1,\bj,\bj'}|d_{\bj,\bj'}\rangle=0 \ ,
\label{e1}
\eeq
and also eigenstates of $h_{2,\bj,\bj'}$ with eigenvalues $K_2$ and 0
\beq
h_{2,\bj,\bj'}|d_{\bj,\bj'}\rangle =K_2|d_{\bj,\bj'}\rangle \ ,
\hskip0.5cm
h_{2,\bj,\bj'}|s_{\bj,\bj'}\rangle =0 \ .
\label{e2}
\eeq
The eigenvalues, which have been plotted in Fig. 1, are
\beq
K_{1}=2
\t^2\sum_{m\ge1}{1\over m!}
{(2 g^2/\omega_0^2)^m \over z-2\epsilon+2\gamma -\omega_0 m} \ ,
\hskip0.5cm
K_{2}=2
\t^2\sum_{m\ge1}{1\over m!}
{(-2 g^2/\omega_0^2)^m \over z-2\epsilon -\omega_0 m} \ .
\label{eigenvalues}
\eeq
$h_{k,\bj,\bj'}$ ($k=1,2$) do not connect different dimer components:
\beq
h_{k,\bj,\bj'}|\d\u,\sigma\rangle=h_{k,\bj,\bj'}|\sigma,\d\u\rangle
=h_{k,\bj,\bj'}|0,\sigma\rangle=h_{k,\bj,\bj'}|\sigma,0\rangle
=0
\hskip0.5cm
(\sigma=\d,\u ;\ \ k=1,2) \ .
\label{different}
\eeq
Moreover, $h_{t,\bj,\bj'}$ switches between dimers with rate $2\tau$:
\[
h_{t,\bj,\bj'}|s_{\bj,\bj'}\rangle =2\tau |d_{\bj,\bj'}\rangle \ ,
\hskip0.5cm
h_{t,\bj,\bj'}|d_{\bj,\bj'}\rangle =2\tau |s_{\bj,\bj'}\rangle \ .
\]
Since always $K_1<K_2$ (cf. Fig. 1), the formation of dimers $|s_{\bj,\bj'}\rangle$ 
is favored. However, when $\gamma<0$ the disadvantage of the dimer $|d_{\bj,\bj'}\rangle$
can be compensated by the local interaction term in $H_0$ which is diagonal and counts
the number of doubly occupied wells. For the state 
$a|s_{\bj,\bj'}\rangle +b|d_{\bj,\bj'}\rangle$ we get the relation
\beq
H_0(a|s_{\bj,\bj'}\rangle +b|d_{\bj,\bj'}\rangle)
=(\epsilon-\gamma + K_1+2\tau b/a)a|s_{\bj,\bj'}\rangle
+(\epsilon+K_2+2\tau a/b)b |d_{\bj,\bj'}\rangle \ .
\label{eigen2}
\eeq
This state is eigenstate of $h_{\bj,\bj'}$ for $\gamma=K_1-K_2+2\tau(b^2-a^2)/ab$ with 
eigenvalues
$\epsilon-\gamma +K_1+2\tau b/a$. For a given $\gamma$ we can evaluate $a$ and $b$ and then determine the
related eigenvalues. In particular, for $\gamma=K_1-K_2$ there are solutions
$a=\pm b$ with corresponding eigenvalues $\epsilon + K_2\pm 2\tau$. 

$K_1$, $K_2$ still depend on the parameter $z$ which has to be determined as a pole  
of the resolvent $G_0(z)$:
\beq
z = \epsilon + K_2(z)\pm 2\tau\ ,
\label{poles}
\eeq
where the lowest pole is the most relevant contribution.
Other solutions of this equation refer to excited states that are not relevant
for the discussion here.
\begin{figure}
\centering
\includegraphics[width=0.7\textwidth]{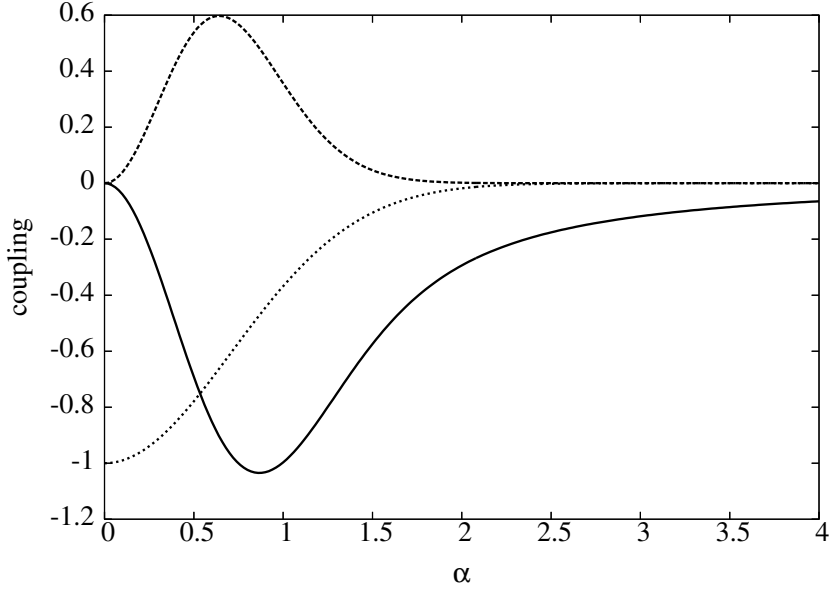}
\caption{Behavior of $K_1$ (full curve), $K_2$ (dashed curve) and $-\tau$ (dotted curve) vs. 
the strength of the fermion-phonon coupling $\alpha =g/\omega_0$. 
The other parameters in $K_1$ and $K_2$ are $\omega_0=J=1$, $\gamma=0$, and $z$ is given by the lowest
pole. All energies are measured in units of $\omega_0$.}
\label{plot2}
\end{figure}

\subsection{Eigenstates for the lattice}

On the lattice with $2M$ wells the effective Hamiltonian $H_0$ can be 
constructed from the effective Hamiltonian in Eq. (\ref{hjj})
by a summation over nearest neighbors.
In order to study simple eigenstates of the lattice system first, the tunneling
of single fermions is excluded from the discussion:
\beq
H_0'=  H_0-\sum_{<\bj,\bj'>}h_{t,\bj,\bj'} \ .
\label{h0'}
\eeq
Eigenstates of $H_0'$ are product dimer states
\beq 
|\Psi_1\rangle=\prod_{<\bj,\bj'>}|s_{\bj,\bj'}\rangle \ , 
\hskip0.5cm
|\Psi_2\rangle=\prod_{<\bj,\bj'>}|d_{\bj,\bj'}\rangle
\eeq 
with eigenvalues
\beq
\lambda_1=(\epsilon-\gamma)M + 2M
\t^2\sum_{m\ge1}{1\over m!}
{(2 g^2/\omega_0^2)^m \over z-(\epsilon-\gamma)M -\omega_0 m} \ ,
\label{eigenvalue1}
\eeq
and
\beq
\lambda_2=\epsilon M + 2M\t^2\sum_{m\ge1}{1\over m!}
{(-2 g^2/\omega_0^2)^m \over z-\epsilon M -\omega_0 m} \ .
\label{eigenvalue2}
\eeq
The alternating dimer state
\[
|\Psi_{AD}\rangle =\prod_{<\bj,\bj'>\in C}\left(
a_{\bj,\bj'}|s_{\bj,\bj'}\rangle +b_{\bj,\bj'}|d_{\bj,\bj'}\rangle
\right)
\]
is also eigenstate of $H_0'$ with eigenvalue
\beq
\lambda_{AD}=(\epsilon-\gamma/2) M  
+ M\t^2\sum_{m\ge1}{1\over m!}
{(2 g^2/\omega_0^2)^m \over z-(\epsilon-\gamma/2) M -\omega_0 m} 
+ M\t^2\sum_{m\ge1}{1\over m!}
{(-2 g^2/\omega_0^2)^m \over z-(\epsilon-\gamma/2) M -\omega_0 m} \ .
\label{eigenvalue3}
\eeq
The coefficients $a_{\bj,\bj'}$ and $b_{\bj,\bj'}$ are either 0 or 1, alternating on the 
lattice. $C$ covers the lattice with dimers, separated by nearest-neigbor bonds (cf. Fig. 2).

Next, the energy $z$ must be determined by the lowest poles of the
resolvent $G_0$. This gives, up to small corrections, for $\lambda_1$ the value 
$z=(\epsilon-\gamma) M$ such that
\beq
\lambda_1/M = 
\epsilon-\gamma - 2{\tau^2\over \omega_0} F(g^2/\omega_0^2)
\hskip0.5cm
{\rm with} \ \ F(x)= \sum_{m\ge1}{(2x)^m\over m!m} \ .
\label{eigenvalue1b}
\eeq  
For $\lambda_2$ the value $z=\epsilon M$ gives
\beq
\lambda_2/M =\epsilon - 2{\tau^2\over \omega_0}F(-g^2/\omega_0^2)
\ ,
\label{eigenvalue2b}
\eeq
and for $\lambda_{AD}$ the value $z=(\epsilon-\gamma/2) M$ leads to
\beq
\lambda_{AD}/M=\epsilon-\gamma/2 - {\tau^2\over\omega_0}F(g^2/\omega_0^2)
- {\tau^2\over\omega_0}F(-g^2/\omega_0^2) 
\ .
\label{eigenvalue3b}
\eeq
Dimer states in a bipartite lattice are not expected to be ground states.
For instance, the N\'eel state gives for $z= (\epsilon -\gamma) M$
\[
\langle\Psi_{N}|H_0|\Psi_{N}\rangle/M =(\epsilon-\gamma)
+\langle\Psi_{N}|\sum_{<\bj,\bj'>}h_{1,\bj,\bj'}|\Psi_{N}\rangle/M 
=(\epsilon-\gamma) - {z_L\tau^2\over \omega_0} F(g^2/\omega_0^2) \ ,
\]
where $z_L$ is the number of nearest-neighbor wells. The corresponding matrix element
for a density-wave state of fermion pairs is $\epsilon$, which is very close to the 
eigenvalue $\lambda_2/M$ of the paired fermion dimers $|\Psi_2\rangle$. 

The single-fermion tunneling has not been included, since the renormalized tunneling rate 
$\tau$ is very small for sufficiently large coupling $g/\omega_0$. 
Therefore, its effect on the eigenvalues will be neglegible, provided the
system is not at a degenerate point. Although the dimer states are not ground states,
the system can stay for very long times in them, in terms of a typical experiment.  
\begin{figure}
\centering
\includegraphics[width=0.5\textwidth]{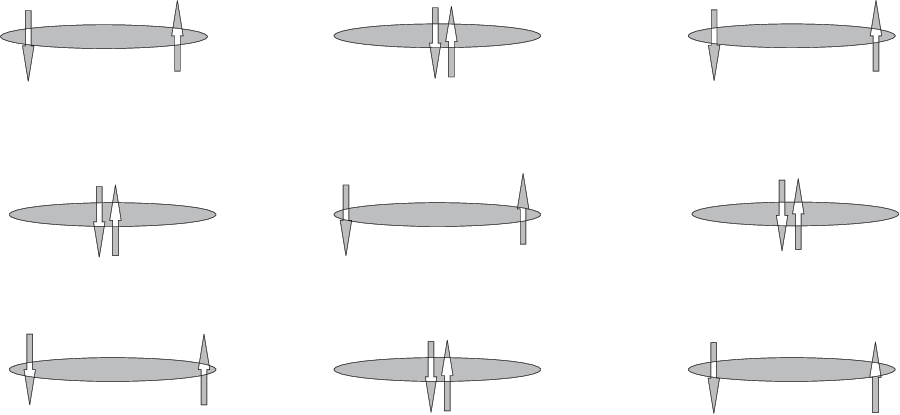}
\caption{The alternating dimer state of Eq. (\ref{dimers}) on a square lattice, 
where the elements represent
$|\d,\u\rangle-|\u,\d\rangle$ and $|\d\u,0\rangle+|0,\d\u\rangle$, respectively.
}
\label{dimer1}
\end{figure}

\section{Discussion and Conclusion}

An effective Hamiltonian $H_0$ has been derived within a truncated recursive projection
approach for a half-filled optical lattice. The truncation should be valid under the assumption 
that the tunneling rate $J$ is small in comparison with the local potential of the HA 
(i.e. for $J\ll\omega_0$). The properties of the effective Hamiltonian in Eq. (\ref{h00}) 
can be summarized as follows: (I) $H_0$ acts in a four-dimensional space at
each well of the optical lattice, spanned by $\{|0\rangle,|\u\rangle,|\d\rangle,|\d\u\rangle\}$.
(II) $H_0$ describes one and two fermion processes between nearest-neighbor wells.
Phonons (i.e. excitations of the HA) are created by the unitary operator $T$ of 
Eq. (\ref{unitary}) and appear in the eigenvalues of the effective Hamiltonian $H_0$ 
as excitation of $m$ phonons. In the case of a system with two wells the eigenstates of 
$H_0$ are superpositions of dimer states, where either the fermions sit in separate wells
or as a pair in only one well. The contribution of each of the two dimers to the eigenstate
can be controlled by the parameter $\gamma=U-2g^2/\omega_0$.  This result was extended to
an optical lattice with $2M$ wells. If single-fermion tunneling is neglected in $H_0$
(this is a good approximation for sufficiently strong fermion-phonon coupling $g/\omega$), dimer
states are eigenstates of the effective Hamiltonian.
Although the dimer states are not ground states of the system, they can be prepared by 
using a modulated optical lattice, where the tunneling barrier is higher between the 
dimers. This is possible because we have control of the optical
lattice (Laser frequency and amplitude) and of the fermion-fermion interaction
$U$ through the Feshbach resonance. After the preparation of the dimer state,
the modulation can be turned off without destroying the dimers, since it is (almost)
an eigenstate. The main effect of the (weak) single-fermion tunneling term 
is to flip singlett dimers into dimers of paired fermions (and vice versa) with a rate $2\tau$.
Therefore, single-fermion tunneling favors the formation of neighboring pairs of different dimers. 
Three different dimer states were studied: states with singlett dimers, states with dimers of paired fermions 
and states with both dimers, placed in alternating order in the lattice. Their corresponding energies 
(given by their eigenvalues with respect to the effective Hamiltonian) are different.
However, there is a point of degeneracy for
\[
\gamma_c = 2{\tau^2\over\omega_0}\left[
F(-g^2/\omega_0^2)-F(g^2/\omega_0^2)]
\right] \ ,
\]
as plotted in Fig. 3.

\begin{figure}
\centering
\includegraphics[width=0.7\textwidth]{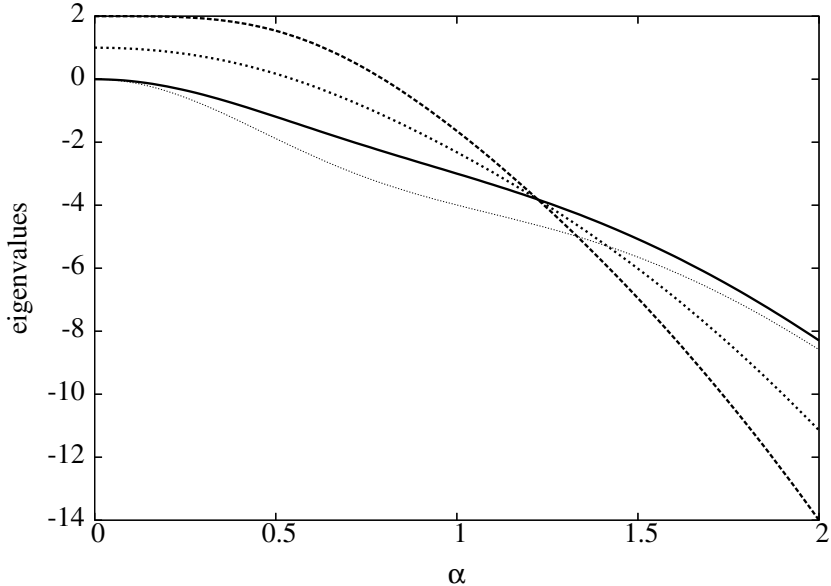}
\caption{Eigenvalues of $H_0'$ of Eq. (\ref{h0'}) for dimer states: $\lambda_1/M$ (full curve) for singletts, 
$\lambda_2/M$ (dashed curve) for paired fermion dimers, $\lambda_{AD}/M$ (dotted curve) for an alternating dimer
state vs. the strength of the fermion-phonon coupling $\alpha =g/\omega_0$ and fixed $J=1$, $U=2$.
For comparison, the diagonal matrix element with respect to the N\'eel state is shown for $z_L=4$ by the thin curve.
All energies are measured in units of $\omega_0$.}
\label{plot2a}
\end{figure}

The interaction between LFA and HA in an optical lattice
can be understood in a simple physical picture. Already in the absence of tunneling, the local
repulsive interaction $U$ of the LFA is affected by the coupling to the HA. It results in a
reduction of $U$ to
$\gamma=U-2g^2/\omega_0$. The reason is that the HA absorb interaction energy from the LFA
by going into an excited state (cf. Eq. (\ref{energy0})). In the presence of a tunneling rate 
$J$ of the LFA
also kinetic energy is taken away by the HA, leading to a reduction of the tunneling rate to
$\tau=J\exp(-g^2/\omega_0^2)$. The renormalization of the tunneling rate and the interaction
strength depends on the parameter $g/\omega_0$. This means that the effect of the HA on 
the LFA is the stronger
the easier it is to excite the HA with energy $\omega_0$. The latter can be controlled in
an optical lattice by tuning the shape of the potential at the bottom of the wells. Since
excitations of energy $\omega_0$ are essential, a deviation of the potential from the assumed harmonic
form at higher energies does not affect the qualitative picture. The fact that $\gamma$ can
be negative (attractive interaction) implies a transition from a repulsive to an
attractive gas of LFA. At half filling the ground states of the LFA in an unfrustrated lattice is
a N\'eel state in the case of strong repulsion and a density wave of paired fermions  
in the case of strong attraction. 
For $\gamma=0$ there is no effective interaction between the LFA. However, this does not mean that
there is a free Fermi gas because the scattering with the HA still affects the tunneling processes.
Our calculation, based on a small tunneling rate, shows that the LFA form
dimer states for $\gamma\approx0$, where either two fermions on nearest neighbor wells
with opposite spin orientation or a locally bound pair of two fermions and an empty well are
formed.  

The dimer states, as well as the N\'eel and the density-wave state, should be observable
in a Stern-Gerlach experiment \cite{ketterle03}. For this purpose, a mixture of two atomic species, 
constituted by the same number of LFA (e.g. $^6$Li) and HA (e.g. $^{87}$Rb or $^{40}$K) is brought
into an optical lattice, where half filling is maintained at low temperature. Then the 
interaction parameter $\gamma=U-2g^2/\omega_0$ is adjusted, either by tuning $U$ through
a magnetic field near a Feshbach resonance, or by tuning the excitation energy $\omega_0$ of the HA
(i.e. the tightness of the wells in the optical lattice). Moreover, the tunneling
rate $J$ is adjusted by the amplitude of the optical lattice. 
After the creation of the quantum state, the optical lattice is turned off to
allow the gas to expand and to measure the spin projections in a Stern-Gerlach experiment.
The N\'eel state, consisting of individual fermions with alternating spin orientation, would separate
in the inhomogeneous magnetic field into two parts. In contrast, the density wave of paired fermions
and the dimer states would not separate so easily.

In conclusion, a mixture of heavy atoms and light fermionic atoms has been investigated.
Inelastic scattering processes generate an attraction between the light atoms and a strong
reduction of the tunneling rate. Depending on the
total interaction, there is the possibility of a N\'eel state, a density wave of paired fermions,
and different dimer states in an optical lattice.

\begin{acknowledgments}

This research was supported by SFB 484.

\end{acknowledgments}

\end{document}